# Experimental evidence of crystal symmetry protection for the topological nodal line semimetal state in ZrSiS

C. C. Gu[1]†, J. Hu[2,3]†, X. L. Chen[1]†, Z. P. Guo[4]†, B. T. Fu[5,6]†, Y. H. Zhou[1], C. An[1,7], Y. Zhou[1], R. R. Zhang[1], C. Y. Xi[1], Q. Y. Gu[4], C. Park[8], H. Y. Shu[9], W. G. Yang[9,10], L. Pi[1,11], Y. H. Zhang[1,11], Y. G. Yao[5], Z. R. Yang[1,7,11]∗, J. H. Zhou[1]∗, J. Sun[4,11]∗, Z. Q. Mao[2,12]∗, M. L. Tian[1,7,11]∗

[1]*Anhui Province Key Laboratory of Condensed Matter Physics at Extreme Conditions, High Magnetic Field Laboratory, Chinese Academy of Sciences, Hefei, Anhui 230031, People's Republic of China*

[2]*Department of Physics and Engineering Physics, Tulane University, New Orleans, LA 70118, USA*

[3]*Department of Physics, University of Arkansas, Fayetteville, AR, 72701, USA*

[4]*National Laboratory of Solid State Microstructures, School of Physics, Nanjing University, Nanjing, Jiangsu 210093, People's Republic of China*

[5]*Beijing Key Laboratory of Nanophotonics and Ultrafine Optoelectronic Systems, School of Physics, Beijing Institute of Technology, Beijing 100081, China*

[6]*College of Physics and Electronic Engineering, Center for Computational Sciences, Sichuan Normal University, Chengdu 610068, China*

[7]*Insititute of Physical Science and Information Technology, School of Physics and Materials Science, Anhui University, Hefei 230601, China*

[8]*HPCAT, Geophysical Laboratory, Carnegie Institution of Washington, Argonne, Illinois 60439, USA*

[9]*Center for High Pressure Science and Technology Advanced Research, Shanghai 201203, People's Republic of China*

[10]*High Pressure Synergetic Consortium, Geophysical Laboratory, Carnegie Institution of Washington, Argonne, Illinois 60439, USA*

[11]*Collaborative Innovation Center of Advanced Microstructures, Nanjing University, Nanjing, Jiangsu 210093, People's Republic of China*

[12]*Department of Physics, Pennsylvania State University, University Park, PA 16803, USA*

†These authors contributed equally to this work. *e-mail: zryang@issp.ac.cn; jhzhou@hmfl.ac.cn; jiansun@nju.edu.cn; zmao@tulane.edu; tianml@hmfl.ac.cn.

# ABSTRACT

Tunable symmetry breaking plays a crucial role for the manipulation of topological phases of quantum matter. Here, through combined high-pressure magneto-transport measurements, Raman spectroscopy, and X-ray diffraction, we demonstrate a pressure-induced topological phase transition in nodal-line semimetal ZrSiS. Symmetry analysis and first-principles calculations suggest that this pressure-induced topological phase transition may be attributed to weak lattice distortions by non-hydrostatic compression, which breaks some crystal symmetries, such as the mirror and inversion symmetries. This finding provides some experimental evidence for crystal symmetry protection for the topological semimetal state, which is at the heart of topological relativistic fermion physics.

# I. INTRODUCTION

Symmetry-protected linearly dispersed Dirac cones[1-4] lead to the distinct properties in topological materials such as light effective mass, high mobility[5] and non-trivial Berry phase[6]. These gapless Dirac nodes are known to be protected by crystal lattice symmetries[3, 4, 7, 8], but an unambiguous experimental demonstration is yet to be done. High pressure acts as an effective means to break crystal symmetry and thus to study the pressure-induced evolution of topological state and to establish a direct connection between the robustness of topological states and the intactness of crystal symmetry. A number of high pressure studies on three-dimensional topological semimetals have been performed previously[9-16] and pressure-induced topological phase transitions (TPT) have been observed in several systems such as $Cd_3As_2$[9, 10], $ZrTe_5$[11], $TaAs$[12] and $ZrSiS$[16], which are either irrelevant to symmetry breaking or originating from drastic structural transitions[9-12]. So far, direct experimental observation of crystal symmetry protection for the non-trivial band topology is still a challenge.

The recently discovered topological nodal-line semimetal ZrSiS is an ideal material to demonstrate how the topological state sensitively relies on the crystal symmetry protection. ZrSiS belongs to a large topological material family *WHM* (*W* = Zr, Hf, or rare earth elements; *H* = Si, Ge, Sn, Sb; *M* = S, Se, Te)[17-24] and hosts two types of Dirac states, *i.e.,* the nodal-line state doubly protected by the mirror- and inversion- symmetry (See Appendix A) and the two-dimensional (2D) Dirac state protected by the non-symmorphic symmetry[18, 23]. Owing to the relatively light elements, the spin-orbit coupling (SOC)-induced gap along the nodal line is small (~ 20 meV along the Γ-X direction)[18, 19], thus relativistic fermion properties such as large magnetoresistivity (MR), high mobility, and non-trivial Berry phase are well-preserved[22, 25-29]. The 2D non-symmorphic Dirac state hardly contributes to low energy excitations since its Dirac node is far away (~ 0.5 eV) from the Fermi level[18, 19]. Due to no other topologically trivial bands crossing the Fermi level in ZrSiS[18], its transport properties should be dominated only by the nodal line fermions, which are supported by the recent de Haas–van Alphen (dHvA) and high-field Shubnikov-de Haas (SdH) experiments[27, 28].

In this work, we show that the topological nodal line state in ZrSiS is found to be robust against ideal hydrostatic pressure, but is suppressed to a trivial state by a weak structural transition or so-called lattice distortions induced by the inhomogeneous compression in non-uniform hydrostatic pressure. The non-uniform hydrostatic pressure generates the shear stress and pressure gradient in the sample and thus may break some crystal symmetries, such as the mirror and inversion symmetries. Our theoretical analysis suggests that the suppression of non-trivial band topology may be attributed to the breakdown of the crystal symmetries caused by weak structural distortions, rather than an obvious structural phase transition or changes in the Fermi surface's morphology.

## II. EXPERIMENTAL DETAILS & METHODS

The ZrSiS single crystals were prepared by a chemical vapor transport method[27]. The stoichiometric mixture of Zr, Si, and S powder was sealed in a quartz tube with iodine being as transport agent (20 mg/cm$^3$). Plate-like single crystals with metallic luster can be obtained via the vapor transport growth with a temperature gradient from 950 °C to 850 °C. The composition and structure of ZrSiS single crystals were checked by x-ray diffraction and Energy-dispersive x-ray spectrometer.

High-pressure resistivity measurements were conducted in a screw-pressure-type DAC using Daphne 7373 oil as the pressure-transmitting medium. Diamond anvils of 300 μm culets and a T301 stainless-steel gasket covered with a mixture of epoxy and fine cubic boron nitride (c-BN) powder were used for high-pressure transport measurements. The four-probe method was applied in the ab plane of single crystals with typical dimensions of 120 × 80 × 10 μm$^3$. The magnetoresistivity experiments under high pressure were performed on the Cell 3 Water-Cooling Magnet of the China High Magnetic Field Laboratory (CHMFL) in Hefei. The measurements were done using a field-sweeping method at fixed temperature. The maximum magnetic field was 27.5 T along the *c* axis. The standard five-probe method was applied on the ab plane of single crystal ZrSiS with dimensions of 100 × 70 × 10 μm$^3$ for the high-pressure Hall measurements. Pressure was calibrated by using the ruby fluorescence shift at room temperature for all experiments[30].

*In situ* high pressure angle-dispersive synchrotron x-ray diffraction (XRD) were performed by using single-crystalline sample powder at 16-BM-D, HPCAT[31] of Advanced Photon Source of Argonne National Laboratory using a Mao-Bell symmetric DAC with Daphne 7373 oil and neon gas as the pressure transmitting medium. A focused monochromatic X-ray beam (~5 μm in FWHM) with wavelength 0.4133 Å was used for the angle-dispersive diffraction. A Mar345 image plate was used to record 2D diffraction patterns. Refinements of the measured XRD patterns were performed by the GSAS software[32]. Pressure dependence of lattice volume was fitted by the usual Brich-Murnaghan equation of states[33].

Raman scattering experiments were carried out at the China High Magnetic Field Laboratory (CHMFL) in Hefei and the Center for High Pressure Science and Technology Advanced Research (HPSTAR) in Shanghai. The Raman spectrum measurement was performed at room temperature in a BeCu-type ST-DAC using a commercial Renishaw Raman spectroscopy system with a 532 nm laser excitation line. The diamond culet was 300 μm in diameter. Neon gas and Daphne 7373 oil were used as the pressure medium to generate hydrostatic or quasi-hydrostatic condition, respectively[34, 35].

To calculate the physical quantities, such as energy bands, SdH frequencies and Fermi surfaces (Figs. 13-15) under pressures, we performed structural optimizations and enthalpy calculations using the projector augmented wave (PAW) method in the Vienna ab-initio simulation package (VASP)[36] with the Perdew-Burke-Ernzerhof (PBE)[37] generalized gradient approximation (GGA) exchange-correlation density functional. The DFT-D3 method[38] is used to take Van der Waals correction into account. We set the cut-off energy of the plane wave basis to 350 eV and sampled the Brillouin zones using the Monkhorst-Pack method with a *k*-mesh spacing of 0.03 $\text{Å}^{-1}$. The electronic structure calculations in Figs. 13-14 are calculated using full-potential linearized augmented plane-wave method[39, 40] in the Wien2k program package[41]. In self-consistent calculations, we used a 1000 *k*-point mesh to sample the Brillouin zone and -7 for the plane-wave vector cut-off parameter $R_{MT} K_{max}$, where $R_{MT}$ is the minimum muffin-tin radius and $K_{max}$ is the plane-wave vector cut-off parameter. We used a second-variation method[42] to take SOC into account. For Fig. 15, we calculated the Fermi surfaces by using a more refined k-point mesh of 62×62×27. We applied the SKEAF interface[43] to calculate the SdH frequencies.

For Figs. 5-6, we used the aforementioned PAW method in the VASP and utilized the optB88-vdW functional to take Van der Waals correction into account. We set the cut-off energy of the plane wave basis to 400 eV and sampled the Brillouin zone using the Monkhorst-Pack method with a *k*-mesh spacing of 0.024 $\text{Å}^{-1}$. The topological invariant $Z_2$ was obtained via the Fu-Kane parity criterion.

### III. ELECTRONIC TRANSPORT

In Fig. 1a, we present the normalized in-plane MR, defined as $\frac{\rho_{xx}(B)-\rho_{xx}(B=0)}{\rho_{xx}(B=0)}$, measured at 1.8 K under different pressures using a diamond anvil cell with Daphne 7373 as the pressure medium. Significant SdH oscillations are observed in the pressure range of 0.5 – 7.4 GPa, where both the amplitude of the oscillations and the MR values are nearly insensitive to the pressure. With increasing pressure above 7.4 GPa, the MR starts to decrease steeply, almost by three orders of magnitude at 20.3 GPa (Fig. 1b), which is accompanied by an abrupt increase in the residual resistivity $\rho_{xx}(0)$ (Fig. 1c), where $\rho_{xx}(0)$ equals the $\rho_{xx}$ value at 1.8 K. At $P$ = 0.5 GPa, the MR reaches $3.5 \times 10^4$ % at 27.5 T without a signature of saturation, consistent with the previous studies at ambient pressure[44]. In addition, both the dramatic drop in MR and the significant increase in $\rho_{xx}(0)$ for $P$ > 7.4 GPa suggest a significant mobility reduction under pressure.

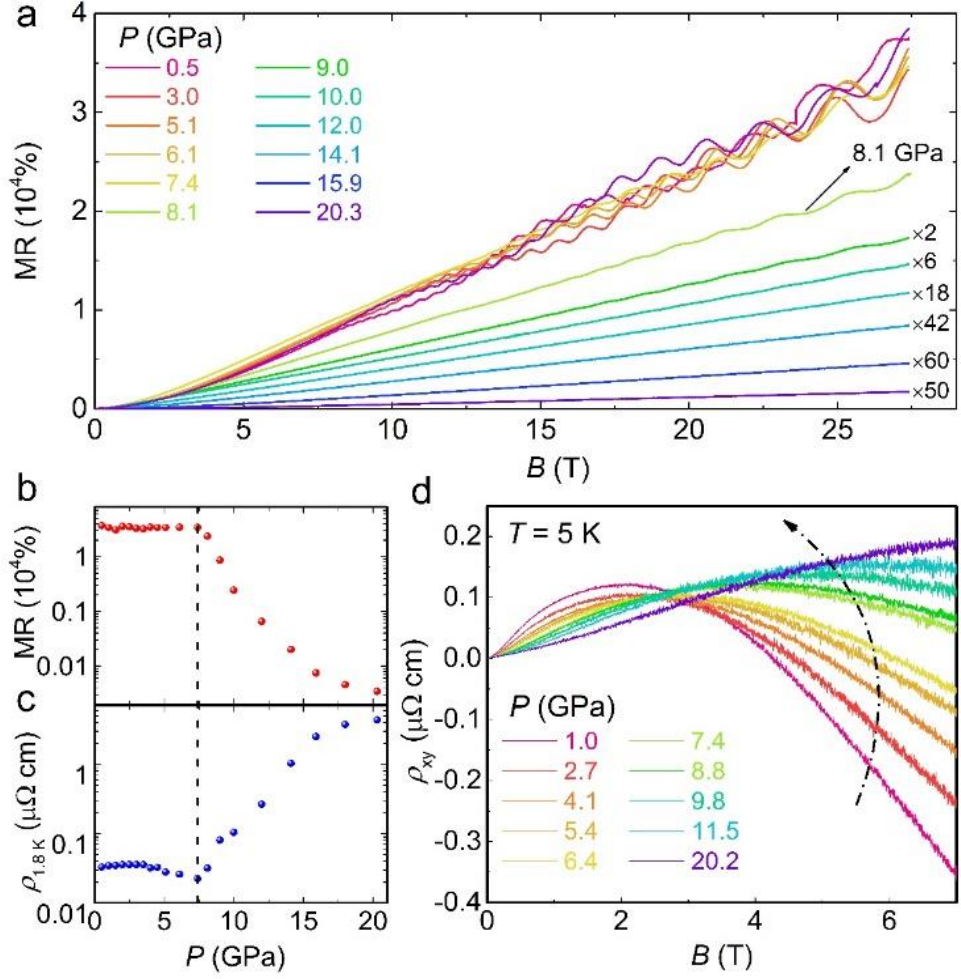

FIG. 1. (a) The evolution of MR for ZrSiS single crystal at selected pressures up to 20.3 GPa and $T$ = 1.8 K. The data at $P \geq 9.0$ GPa were amplified for clarity and the amplification factors were given above the curves. The MR ratio for the highest field 27.5 T (b) and the residual resistivity (c) at 1.8 K are presented as a function of pressure, respectively. (d) Field dependences of Hall resistivity $\rho_{xy}$ at various pressures from 1.0 to 20.2 GPa.

The drastic mobility drop caused by pressure is confirmed by our pressure dependent Hall resistivity $\rho_{xy}(B)$ measurements at $T$ = 5 K, in Fig. 1d. Like the ambient pressure measurements[25, 27], the multiband signature of ZrSiS is clearly manifested in the non-monotonic field dependence of $\rho_{xy}(B)$ with a hump in the low-field range. The hump is broadened and shifted to higher field with increasing pressure, becoming non-observable up to 7 T when $P > 9.8$ GPa. These observations reflect the gradual evolution of electronic band structure with pressure. According to the fit of $\rho_{xy}(B)$ to the classic two-band model[45] (Appendix B), the carrier mobilities ($\mu_e$, $\mu_h$) drop significantly at $P > 5.4$ GPa, which induces the remarkable drop of MR and the sharp increase of residual resistivity above 7.4 GPa. However, the electron and hole densities ($n_e$, $n_h$) do not exhibit prominent changes above 7.4 GPa.

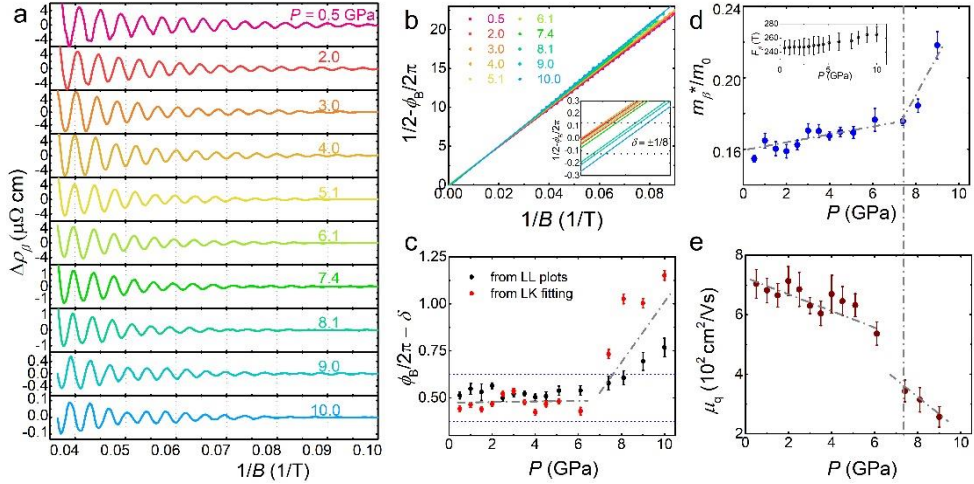

FIG. 2. (a) $\Delta\rho$ for $F_\beta$ component is plotted as a function of $1/B$ at selected pressures with $T = 1.8$ K. (b) The Landau Level fan diagram of $F_\beta$. The inset of (b) is the enlarge view of the intercepts. The evolution of Berry phase (c), effective mass (d) and quantum mobility (e) for $\beta$ pocket upon compression, respectively. The inset of (d) shows the evolution of $F_\beta$ upon compression. The black dash line indicates the critical pressure $P = 7.4$ GPa, where the topological nodal line state is suppressed to a trivial state.

The direct implication of the remarkable decrease of carrier mobility above 7.4 GPa is that the nodal line state of ZrSiS might be suppressed to a topological trivial state. This can be confirmed by the pressure evolution of quantum oscillation properties of ZrSiS. As shown in Fig. 1a, ZrSiS exhibits the SdH oscillations with two major oscillation frequencies ($F_\alpha \approx 24$ T, $F_\beta \approx 246$ T, see Fig. 8) in good agreement with the ambient-pressure studies[25-27, 29, 44] and can be ascribed to the Fermi surface enclosing the Dirac nodal-line[21, 27]. In this work, we will mainly focus on the higher frequency ($F_\beta$) oscillation component (Appendix C), which belongs to the nodal point along the ZR path (Fig. 13a). In Fig. 2a, we present the higher frequency oscillation components at different pressures up to 10.0 GPa, which were obtained by removing the smooth MR background and filtering out the low frequency component. Notably, the oscillation patterns exhibit a significant evolution with increasing pressure. According to the Lifshitz-Kosevich (LK) formula[46, 47], *i.e.*, $\Delta\rho \propto \cos[2\pi(\frac{F}{B} - \frac{1}{2} + \frac{\phi_B}{2\pi} - \delta)]$, where $\phi_B$ is Berry phase and $\delta = \pm 1/8$ is the additional factor due to the three-dimensional Fermi surface, the change in the phase factor should be attributed to the variation of Berry phase. Below 6.1 GPa, a non-trivial Berry phase is obtained with the factor $\frac{\phi_B}{2\pi} - \delta$ scattering around 0.5, as shown in Figs. 2b and 2c. However, further increasing pressure above 7.4 GPa leads to a sharp increase in $\frac{\phi_B}{2\pi} - \delta$, implying an abrupt change of band topology. Coincidently, the effective cyclotron mass $m_\beta^*$ and quantum mobility $\mu_q$ extracted from the temperature and field damping of the oscillation amplitude (see Fig. 12) for the $F_\beta$ band also display striking changes near the critical pressure where the Berry phase becomes trivial, as exhibited in Figs. 2d and 2e. Upon compression up to 7.4 GPa, $m_\beta^*$ suddenly increases while $\mu_q$ drops drastically. Because of the intimate connection between high mobility and topological protection[5], the sharp decrease in quantum mobility implies the loss of topological protection for nodal-line fermions in ZrSiS, which thus results in remarkable changes in transport properties.

### IV. STRUCTURAL PROPERTIES

In order to understand the origin of the pressure induced topological transitions, we have performed systematic high-pressure synchrotron XRD study with Daphne 7373 as pressure medium. As shown in Fig. 3a, the diffraction spectra display smooth evolution without any abrupt changes up to $P$ = 37.9 GPa. All diffraction peaks can be indexed with the *P*4/*nmm* tetragonal structure, indicating the absence of a drastic structural transition as well as the neon gas case in Fig. 3b. We have performed Rietveld refinements to determine lattice parameters at each pressure. The lattice constants (Fig. 3c) and the unit-cell volume (Fig. 3e) vary smoothly with pressure, while the *c/a* axial ratio for Daphne oil exhibits a kink near 6.9-7.8 GPa (Fig. 3d), where the electronic properties display sudden changes as discussed above. Such a feature is reminiscent of the electronic topological transition (ETT) tuned by pressure in layered topological insulators[48-50], which is usually not accompanied by a discontinuity of the volume but by a change in the compressibility[48, 49]. Although ZrSiS undergoes a critical change in the compressibility around 6.9-7.8 GPa, the ETT is not expected to occur, since the carrier densities do not change remarkably above 7.4 GPa (Fig. 7b). Pressure-induced topological transitions in the absence of structural transition in several materials[9, 11, 16, 51] originate from the continuous tuning of electronic band structure rather than being relevant to crystal symmetry. In ZrSiS, however, the SdH oscillation frequency $F_\beta$ varies only by 7.5% from 0.5 to 10.0 GPa (Fig. 2d, the inset), in sharp contrast with the aforementioned remarkable changes in electronic band structure. Since quantum oscillation frequencies are dependent on the extremal cross-section areas of the Fermi surface, the smooth variation of $F_\beta$ across the critical pressure (~7.4 GPa) for ZrSiS indicates that its Fermi surface morphology undergoes only slight changes through the critical pressure and the topological phase transition is unlikely caused by energy band shift (Appendix D).

The high fields transport measurements under high pressures do show that the nodal line state of ZrSiS is suppressed to a topologically trivial state when the pressure is increased above 7.4 GPa. It has been known that the hydrostatic compression range for Daphne 7373 is limited to only 2.3 GPa[34], while the neon case could be at least to 15 GPa[35]. So, the *c/a* anomaly near 6.9-7.8 GPa observed in Daphne 7373 is absent in the neon gas pressure experiment (Fig. 3d), indicating the non-hydrostatic compression in Daphne 7373. In fact, similar phenomena have previously been observed in pressurized Zn[52]. When the pressure transmitting medium, Daphne 7373, starts to solidify, the lattice under non-hydrostatic compression would undergo different compression response along the *c* and *a* axes (Fig. 3c), thus resulting in the *c/a* anomaly. This is also confirmed by the different strain evolutions between the non-hydrostatic (Daphne 7373) and hydrostatic (Neon) compression (Fig. 16).

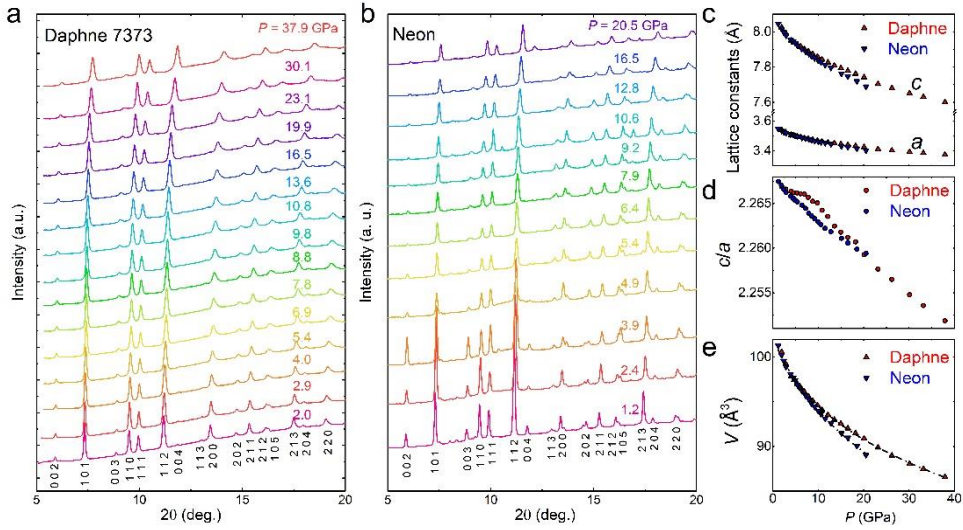

FIG. 3. Room temperature representative diffraction patterns at high pressure (a) with Daphne 7373 from 2.0 to 37.9 GPa; (b) with neon from 1.2 to 20.5 GPa. Pressure dependences of lattice parameters (c), *c/a* ratio (d) and volume (e).

High pressure Raman experiments enables us to further verify the structural distortions produced by non-hydrostatic compression. At ambient pressure, three vibrational modes at 209.7 (denoted as $P_1$), 301.7 ($P_2$), and 310.5 ($P_3$) cm$^{-1}$ are identified in the Raman shift (Figs. 4a and 4b) in the configuration $z(xx)\bar{z}$. $P_1$ and $P_2$ can be attributed to the anti-symmetric vibration modes ($A_{1g}$) of Zr and S atoms, while $P_3$ is due to the vibration of Si atoms along the *c*-axis ($B_{1g}$)[53]. Increasing the pressure leads to remarkable stiffening for all three Raman modes for both the Ne gas and Daphne fluid, as summarized in Figs. 4c and 4d respectively. No additional Raman modes are observed up to 30 GPa. As expected, all three Raman modes display linear pressure dependence for Ne gas up to 20 GPa. In contrast, when using Daphne 7373 as pressure medium, the pressure dependences of the Raman modes show slope changes around 6~8 GPa (black arrows) in Fig. 4d, where the anomalies are unveiled in both magnetotransport and XRD measurements. The absence of anomaly in the neon medium could be attributed to the nearly perfect hydrostatic pressure, whereas the Daphne 7373 medium leads to inhomogeneous compression, thus resulting in slight structural distortions even at relatively low pressures.

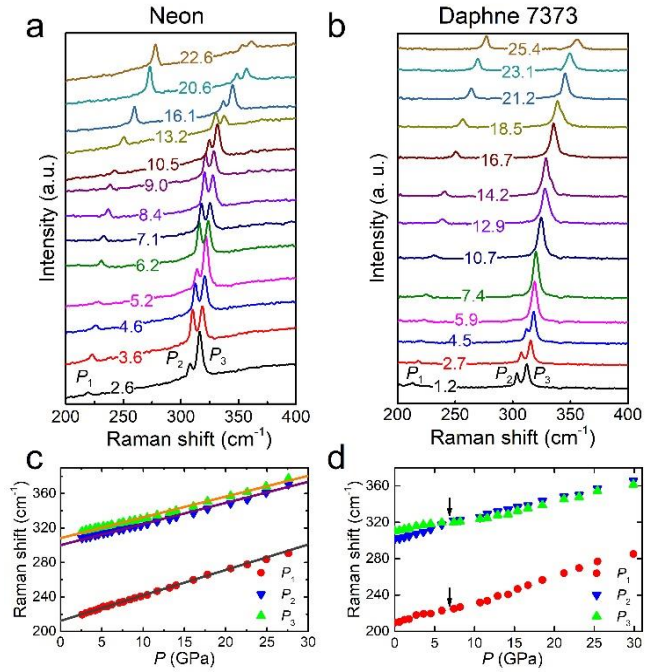

FIG. 4. High-pressure Raman spectra were measured using Neon gas (a) and Daphne 7373 oil (b) as the pressure medium, respectively. The Raman configuration is $z(xx)\bar{z}$. The pressure dependences of corresponding Raman modes are shown in (c, Neon) and (d, Daphne7373), respectively.

## V. RESULTS & DISCUSSIONS

The Dirac nodal lines of ZrSiS are doubly protected by both the mirror symmetry $M_z$ and inversion symmetry $P$ on the $k_z = 0$ and $k_z = \pi$ planes. These symmetries are well maintained under an ideal hydrostatic pressure with neon gas as the pressure transmitting medium. Nevertheless, when Daphne 7373 is used as the pressure transmitting medium, the shear stress as well as the pressure gradient can be generated in the sample space, leading to slight lattice distortions as manifested in the anomalies in the pressure coefficients of Raman frequencies (Fig. 4d), the *c/a* axial ratio (Fig. 3d), and the endured strain (Fig. 16) near 6-8 GPa. Although our XRD and Raman measurements do not show any clear symmetry change, the recent high-pressure structural work of ZrSiS has observed some signatures of weak structural transition[54]. Then the symmetry breaking due to non-hydrostatic compression may take place on some very small scale. It is similar to the case of $Ba(Fe_{1-x}Co_x)_2As_2$[55], in which the $C_4$ symmetry breaking occurs on sub-Ångstrom scale, which cannot be detected by the XRD measurement. The most possible lattice distortion with the lowest energy cost would be the sliding between layers of atoms including the intra- and inter-unit cell parts (Fig. 17), which often happen in layered materials[56]. The lattice distortion within a unit cell only breaks the mirror symmetry and shifts the position of the nodal line in the Brillouin zone but do not open gaps on the nodal line. In Appendix A, we have demonstrated that the breaking of either mirror symmetry or the inversion symmetry does not open a global gap at the nodal line without SOC. To properly model the lattice distortion due to the non-hydrostatic pressure, we mimic the shear stress with a gradient within a 1:1:3 supercell, in which the nodal lines along the MΓ and ΓX paths are downfolded into the reduced Brillouin zone (Figs. 5a-5c). We find that the lattice distortion could open a finite and variable gap along the whole nodal lines, as shown in Fig. 5b. Moreover, when the distortion is enhanced further [e.g. (*α, β, Υ*) = (88, 89, 90) degree and the center of supercell shifts by (0.2, 0.4, 0) Å], the band gaps increase continuously (Fig. 5c). This means that structural distortions created by non-hydrostatic compression lead to the topological transition, which accounts for the disappearance of non-trivial Berry phase as well as a large decrease in mobility upon compression (> 7.4 GPa) in Daphne 7373. In other words, our combined experimental and theoretical efforts unequivocally demonstrate that crystal symmetries play a crucial role in protecting topological electronic states.

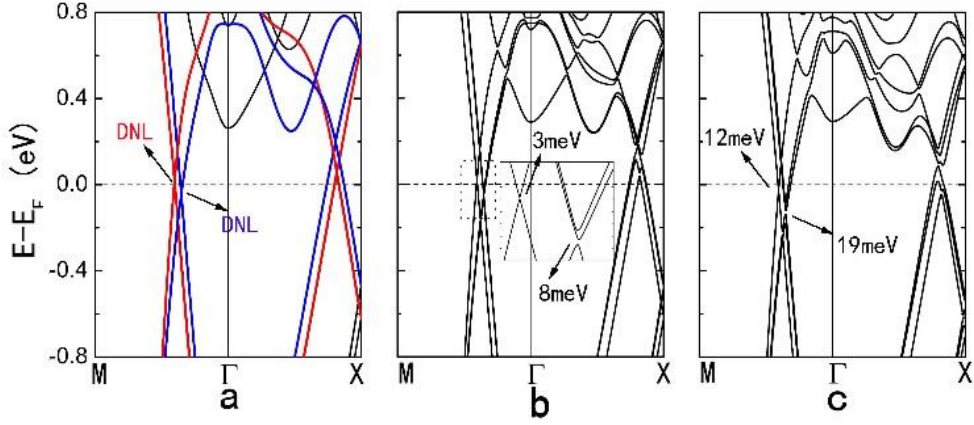

FIG. 5. Band structures of nodal lines along the MΓ and ΓX paths of ZrSiS without SOC in a 1:1:3 supercell shown in Fig. 17d with different lattice distortions: (a) without lattice distortions; (b) with (α, β, ϒ) = (89, 89.5, 90) degree and the shift of the center unit cell parallel to the atom layers by (0.1, 0.2, 0) Å; (c) with (α, β, ϒ) = (88, 89, 90) degree and the shift of the center unit cell parallel to the atom layers by (0.2, 0.4, 0) Å.

## VI. SUMMARY

In summary, we find that the topological nodal-line state of ZrSiS becomes a trivial state when pressure is increased above 7.4 GPa with Daphne 7373 as the pressure transmitting medium. The theoretical analysis suggests that the pressure-induced topological phase transition in ZrSiS can be induced by the inhomogeneous compression, which may break down some crystal symmetries and thus destroy the topological nodal-line state.

**APPENDIX A: Double protection of nodal lines in ZrSiS by both mirror and inversion symmetries**

The layered material ZrSiS has space inversion $P$, time reversal $T$ and mirror symmetry $M_Z$. We notice that the nodal lines are doubly protected by the $M_Z$ and $P$. Here we would like to prove that the breaking of either $M_Z$ and $P$ does not open a gap at the nodal line. In general, a nodal line can be described by a two-band $k.p$ model

$$H(k) = f_0(k)\tau_0 + f_1(k)\tau_x + f_2(k)\tau_y + f_3(k)\tau_z,$$

Where $\tau_{x,y,z}$ are three Pauli matrices. The time reversal operation $T$ reduces to be a complex conjugation operator for the spinless case. Then we consider the constraint of those three symmetries on the $k.p$ model.

$$PH(k)P^{-1} = H(-k);$$
$$TH(k)T^{-1} = H(-k);$$
$$M_z H(k_x, k_y, k_z) M_z^{-1} = H(k_x, k_y, -k_z);$$

Here $P$ and $M_Z$ can be written as $\tau_z$ if the eigenvalues of an operation for two bands have opposite signs, and written as $\tau_0$ if the eigenvalues of an operation have the same signs.

For the case of ZrSiS, we can see that two eigenvalues of $P$ at the $\Gamma$ point is (+-), which indicates $P = \tau_z$. Similarly, we obtain $M_Z = \tau_z$. One can identify the two constraints from both $T$ and $P$:

$$T = K$$
$$f_1(k) = f_1(-k)$$
$$-f_2(k) = f_2(-k)$$
$$f_3(k) = f_3(-k)$$

$$P = \tau_z$$
$$-f_1(k) = f_1(-k)$$
$$-f_2(k) = f_2(-k)$$
$$f_3(k) = f_3(-k)$$

The constraints from *T* and *P* implies that $f_1 = 0$, $f_2$ is an odd function about $k$, and $f_3$ is an even function about $k$.

Let us turn to consider $M_z = \tau_z$ and have the constrains:

$$M_z = \tau_z$$
$$f_2(k_x, k_y, k_z) = -f_2(k_x, k_y, -k_z)$$
$$f_3(k_x, k_y, k_z) = f_3(k_x, k_y, -k_z)$$

Based on the analysis above, one gets the approximate expressions of $f_2$ and $f_3$:

$$f_2 = t_1 k_z + t_2 k_z k_x k_y + t_3 k_z^3 + o(k^3)$$
$$f_3 = m_0 + m_1 k_x^2 + m_2 k_y^2 + m_3 k_z^2 + o(k^2)$$

where $m_i, t_i$ are the material-dependent parameters that can be determined by first principles calculations. Thus, we can write the Hamiltonian as follows:

$$H = (t_1 k_z + t_2 k_z k_x k_y + t_3 k_z^3) \tau_y + (m_0 + m_1 k_x^2 + m_2 k_y^2 + m_3 k_z^2) \tau_z$$

Retaining the terms up to the second order in $k_\alpha$ ($\alpha = x, y, z$) leads to:

$$H = (t_1 k_z) \tau_y + (m_0 + m_1 k_x^2 + m_2 k_y^2 + m_3 k_z^2) \tau_z.$$

Note that this Hamiltonian describes a nodal line at $k_z = 0$ plane for $m_0 > 0$, and $m_1, m_2, m_3 < 0$.

Now let us consider a small perturbation that breaks the $M_z$ symmetry but respects *P* symmetry. The corresponding $f_2$ can be written as

$$f_2 = t_1 k_z + t_2 k_x + t_3 k_y; t_1 \gg t_2, t_1 \gg t_3,$$

where the last two terms in $f_2$ originate from the small perturbation. The Hamiltonian turns out to be:

$$H = (t_1 k_z + t_2 k_x + t_3 k_y) \tau_y + (m_0 + m_1 k_x^2 + m_2 k_y^2 + m_3 k_z^2) \tau_z.$$

It is clear that $f_2 = 0$ gives rise to a slant plane, while $f_3 = 0$ leads to an ellipsoidal surface. The plane intersects with the ellipsoidal surface and will give us a nodal line located at the slant plane (not at $k_z=0$). Therefore, we have shown that, for a system with both $P$ and $M_z$, the nodal line is doubly protected. The breaking of $M_z$ alone does not open a gap at the nodal line but only shifts the position of the nodal line in the Brillouin zone. Note that the discussion of breaking inversion symmetry $P$ is similar.

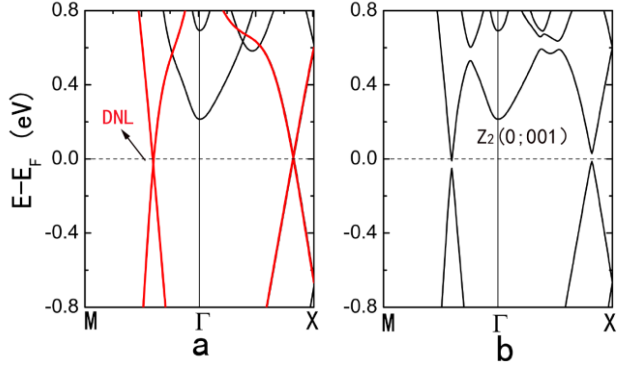

FIG. 6. Band structures along MΓX direction without and with SOC. (a) without SOC. (b) SOC opens a finite gap at the Dirac nodal line (DNL) and ZrSiS becomes a weak topological insulator with topological index $Z_2 = (0;001)$.

**APPENDIX B: The evolution of carrier densities and mobilities upon compression**

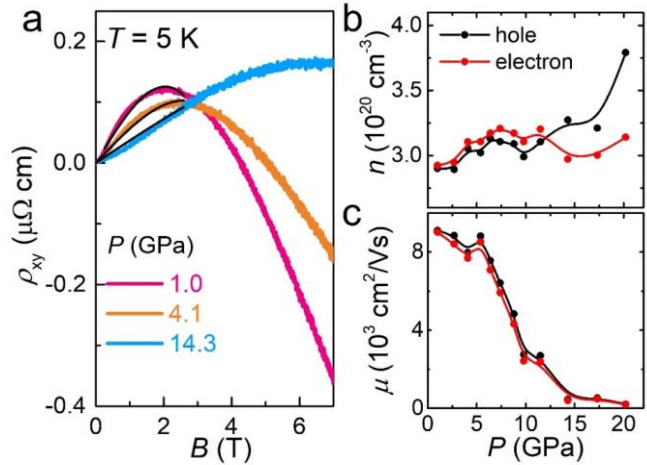

FIG. 7. The two-band model fitting of Hall resistivity of ZrSiS. (a) the fitting curves of the Hall resistivity with two-band mode. The Black lines are the fitting results. Pressure dependences of carrier concentrations (b) and mobilities (c). Black: hole; red: electron.

By carefully fitting the hall resistivity $\rho_{xy}(B)$ to this two-band model (Fig. 7a),

$$\rho_{xy}(B) = \frac{B}{e} \frac{(n_h\mu_h^2 - n_e\mu_e^2) + (n_h - n_e)\mu_e^2\mu_h^2 B^2}{(n_h\mu_h + n_e\mu_e)^2 + [(n_h - n_e)\mu_e\mu_h B]^2} \qquad (1)$$

where $n$ and $\mu$ are the charged carrier (electron $e$ or hole $h$) density and mobility, respectively. The pressure dependences of carrier densities and mobilities are derived, as displayed in Figs. 7b and 7c.

**APPENDIX C: The details about the quantum oscillations of ZrSiS under different pressures**

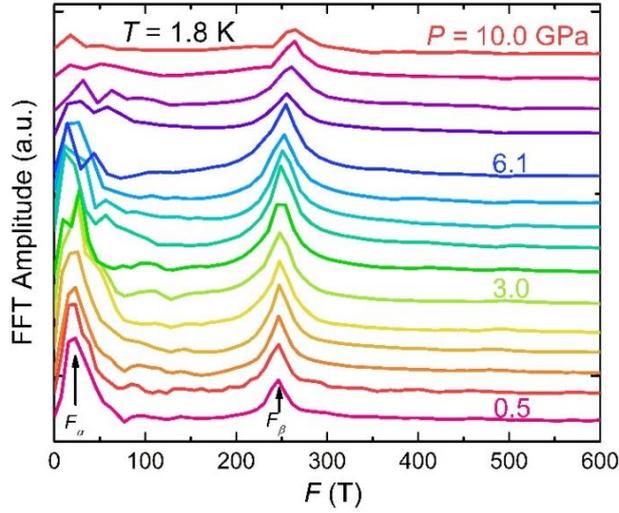

FIG. 8. The FFT amplitude spectra of ZrSiS. FFT spectra at selected pressures from 0.5 to 10.0 GPa. The black lines indicate the two major frequencies $F_\alpha$ = 24 T and $F_\beta$ = 246 T at $P$ = 0.5 GPa.

Figure 8 gives the Fast Fourier Transformation (FFT) amplitude spectra of ZrSiS. The raw data of the oscillation part under selected pressures are tracked by subtracting the smooth magnetoresistivity background from the $\rho_{xx}(B)$ curves, which is shown in Fig. 9. The resistivity oscillations are gradually suppressed upon compression, becoming hardly observable above 10.0 GPa. Analysis for the lower frequency ($F\alpha$) band is very difficult, due to too few oscillation peaks observed in the measured field range and strong Zeeman splitting, as shown in Fig. 10.

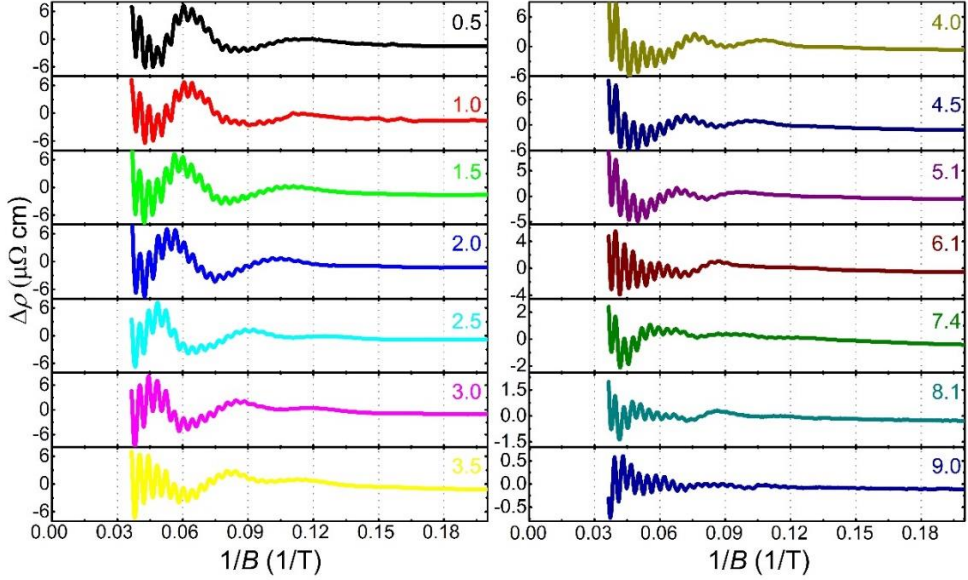

FIG. 9. The SdH oscillation components of magnetoresistivity in ZrSiS. The numbers represent pressures in unit of GPa. The data are obtained by removing the smooth magnetoresistivity background.

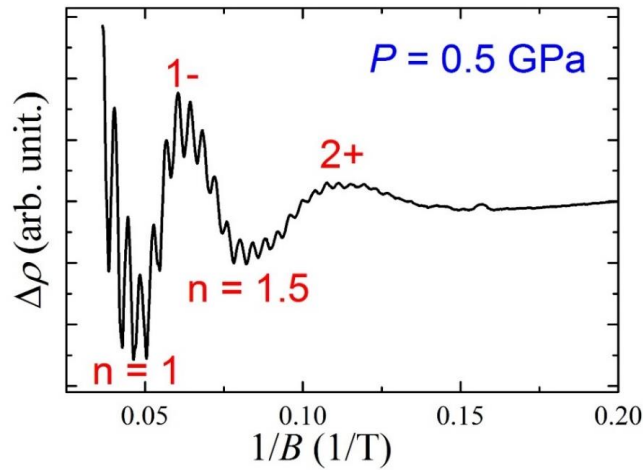

FIG. 10. The SdH oscillation components at 0.5 GPa. In the whole measured field range, there is only one distinguishable period accompanied with apparently Zeeman

splitting, which is insufficient to extract the Berry phase and effective mass of the $\alpha$ pocket from the data.

According to the Lifshitz-Kosevich (LK) formula,

$$\Delta\rho \propto R_T R_D \cos[2\pi(\frac{F}{B} - \frac{1}{2} + \frac{\phi_B}{2\pi} - \delta)]$$

$$R_T = \alpha m^* T/[B\sinh(\alpha m^* T/B)]$$

$$R_D = \exp(-\alpha m^* T_D/B)$$

as well as,

$$\mu_q = \beta/(m^* T_D)$$

where $\alpha$ and $\beta$ are constants, which are equal to 14.69 T/K and 0.21 m$^2$K/Vs, respectively, $B$ is the magnetic flux density and $m^* = m/m_e$. The information on the Berry phase $\frac{\phi_B}{2\pi} - \delta$, effective cyclotron mass $m_\beta^*$ and quantum mobility $\mu_q$ are extracted by fitted the oscillation part with the LK formula, as depicted in Figs. 11-12. Due to $\rho_{xx} \gg \rho_{xy}$ in the whole measured field range, the change of Hall resistivity would not affect the determination of Berry phase simply using $\rho_{xx}$.

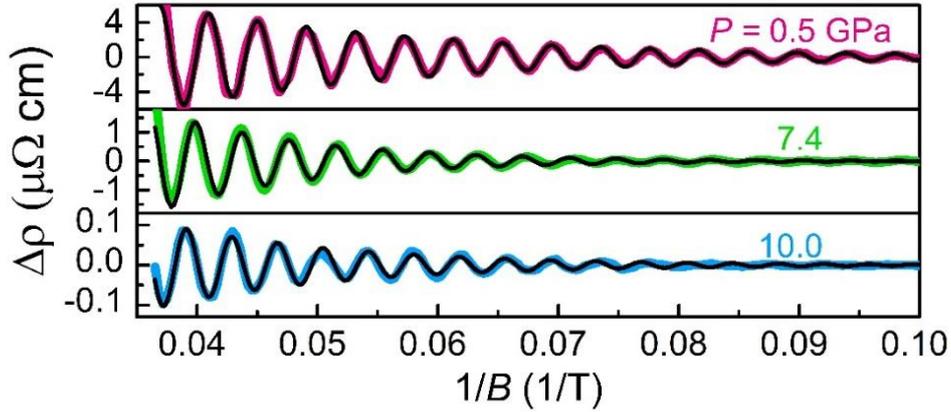

FIG. 11. The fit for the Berry phase of $F_\beta$. Based on the LK formula, the Berry phase at selected pressures were obtained. The black lines are the fitted result.

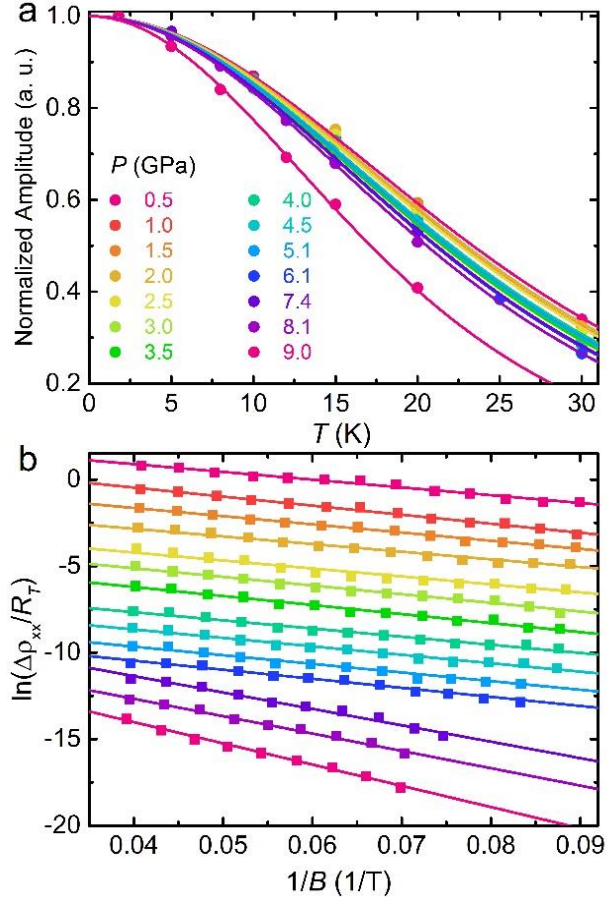

FIG. 12. (a) The fits of the FFT amplitudes of $F_\beta$ to the temperature damping term $R_T$ with the LK formula from 0.5 to 9.0 GPa. (b) Dingle plots for the $F_\beta$ at selected pressures. The data were shifted in the y-axis for clarity.

**APPENDIX D: The calculated band structure of ZrSiS under pressure**

As shown in Figs 13-14, the hydrostatic pressure up to 20 GPa only leads to slight shifts in the energy of the Dirac line nodes, whereas the sharply dispersed Dirac bands remain intact. The non-symmorphic symmetry protected Dirac node below Fermi level at the X point remains nearly unaffected as well.

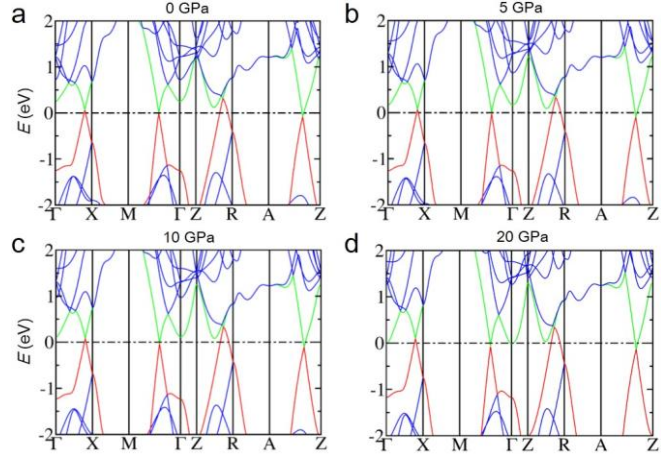

FIG. 13. Calculated band structure calculation without SOC in different pressures. (a) $P = 0$ GPa; (b) $P = 5$ GPa; (c) $P = 10$ GPa; (d) $P = 20$ GPa.

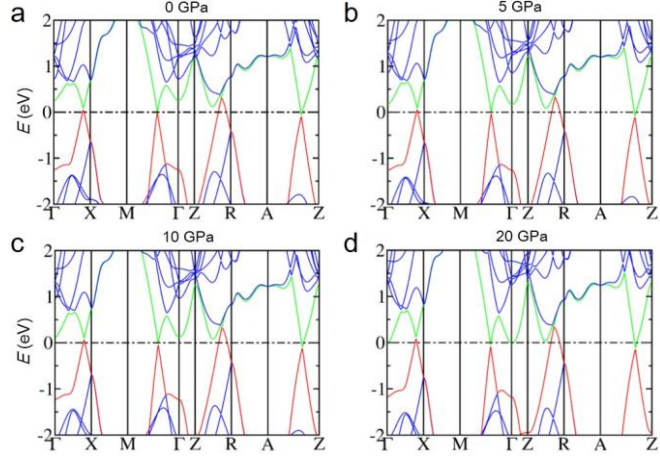

FIG. 14. Calculated band structure calculation with SOC in different pressures. (a) $P = 0$ GPa; (b) $P = 5$ GPa; (c) $P = 10$ GPa; (d) $P = 20$ GPa.

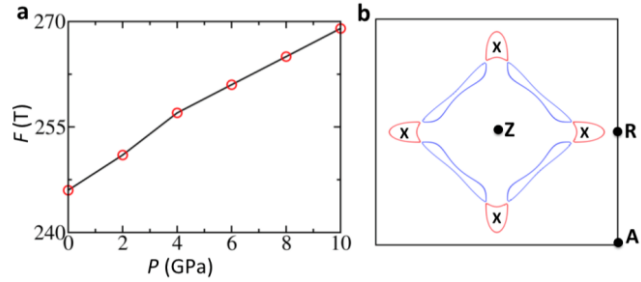

FIG. 15. (a) Calculated SdH oscillation frequencies corresponding to extremal orbits $F_\beta$. (b) Fermi surfaces at $k_z = \pi$ plane in ambient condition, the red lines are extremal

orbits $F_\beta$ observed in experiments, and the crosses inside the extremal orbits mark the locations of the crossing points of nodal lines in $k_z = \pi$ and nodal lines along $k_z$ direction.

**APPENDIX E: The analysis of FWHM of the peaks in XRD spectra.**

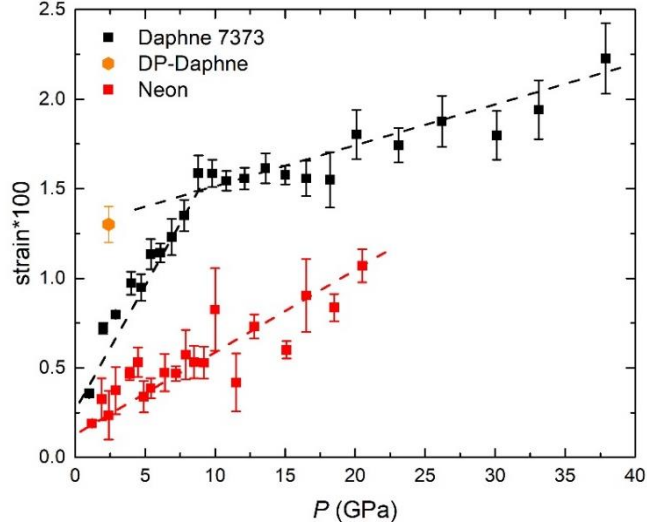

FIG. 16. Pressure dependence of the endured strain in ZrSiS under compression. The black (red) curve represents the evolution of strain on the sample induced by pressure transmitting medium Daphne 7373 (Neon).

The strain dependence of diffraction line widths could be described by the following relation[57, 58]:

$$\text{FWHM}^2 \cos^2\theta = (\lambda/d)^2 + \sigma^2 \sin^2\theta,$$

where FWHM is the full-width at half-maximum of the diffraction profile on $2\theta$-scale. The symbols d, $\lambda$, and $\sigma$ denote the grain size, X-ray wavelength, and deviatoric strain, respectively. Upon compression, the strain on the sample induced by Daphne 7373 increases much more quickly than that in Neon. Moreover, the former shows a remarkably reduced slope above ~ 8 GPa, indicating the stress release by a slight lattice distortion. The discrepancy between the two pressure transmitting mediums demonstrates the existence of pressure gradient and shear stress upon non-hydrostatic compression (Daphne 7373).

**APPENDIX F: A sketch about the evolution of the crystal lattice under hydrostatic/non-hydrostatic compression.**

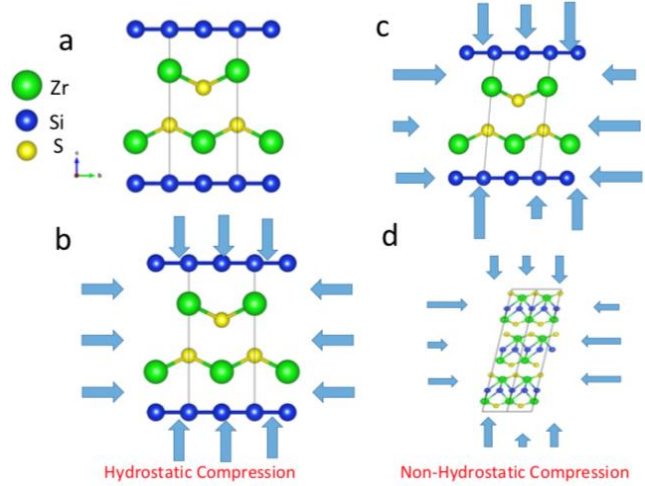

FIG. 17. An illustration for the evolution of the crystal lattice under hydrostatic/non-hydrostatic compression. (a) The unit cell of ZrSiS with no any compression. (b) The unit cell under hydrostatic compression. (c) The unit cell under non-hydrostatic compression. (d) The 1:1:3 supercell under non-hydrostatic compression. The 1:1:3 supercell consists of one primitive cell along the $x$ or $y$ direction but three primitive cells along the $z$ direction. The lattice distortions due to the non-hydrostatic pressure essentially consist of the intra-unit cell part and the inter-unit cell part in the layered ZrSiS such that both the mirror and inversion symmetries are broken simultaneously. Different choices of such lattice distortions only change the magnitudes of gaps at the nodal line.